# IONIZATION OF POLYCYCLIC AROMATIC HYDROCARBON MOLECULES AROUND THE HERBIG AE/BE ENVIRONMENT[*]

ITSUKI SAKON[†]

*Department of Astronomy, Schools of Science, University of Tokyo,*
*7-3-1 Hongo, Bunkyo-ku, Tokyo 113-0033, Japan*

YOSHIKO K. OKAMOTO

*Institute of Astrophysics and Planetary Sciences, Ibaraki University, Japan*
*2-1-1 Bunkyo, Mito 310-8512, Ibaraki, Japan*

HIROKAZU KATAZA

*Institute of Space and Astronautical Science, Japan Aerospace Exploration Agency,*
*3-1-1 Yoshinodai, Sagamihara, Kanagawa 229-8510, Japan*

TAKASHI ONAKA

*Department of Astronomy, Schools of Science, University of Tokyo,*
*7-3-1 Hongo, Bunkyo-ku, Tokyo 113-0033, Japan*

HIDEHIRO KANEDA, MITSUHIKO HONDA[#]

*Institute of Space and Astronautical Science, Japan Aerospace Exploration Agency,*
*3-1-1 Yoshinodai, Sagamihara, Kanagawa 229-8510, Japan*

We present the results of mid-infrared N-band spectroscopy of the Herbig Ae/Be system MWC1080 using the Cooled Mid-Infrared Camera and Spectrometer (COMICS) on board the 8m Subaru Telescope. The MWC1080 has a geometry such that the diffuse nebulous structures surround the central Herbig B0 type star. We focus on the properties of polycyclic aromatic hydrocarbons (PAHs) and PAH-like species, which are thought to be the carriers of the unidentified infrared (UIR) bands in such environments. A series of UIR bands at 8.6, 11.0, 11.2, and 12.7μm is detected throughout the system and we find a clear increase in the UIR 11.0μm/11.2μm ratio in the vicinity of the central star. Since the UIR 11.0μm feature is attributed to a solo-CH out-of-plane wagging mode of cationic PAHs while the UIR 11.2 μm feature to a solo-CH out-of-plane bending mode of neutral

---

[*] This work is *based on data collected at Subaru Telescope, which is operated by the National Astronomical Observatory of Japan.*
[†] I.S. is financially supported by the Japan Society for the Promotion of Science (JSPS)
[#] Current address: Department of Information Science, Kanagawa University, 2946 Tsuchiya, Hiratsuka, Kanagawa, 259-1205, Japan

PAHs, the large 11.0μm/11.2μm ratio directly indicates a promotion of the ionization of PAHs near the central star.

## 1. Introduction

The unidentified infrared (UIR) bands are a series of emission bands observed at 3.3, 3.4, 6.2, 7.7, 8.6, 11.0, 11.2, 12.0, and 12.7 μm together with some other fainter features. They have been ubiquitously observed in various astrophysical objects, including reflection nebulae (RNe), HII regions, planetary nebulae (PNe), post-AGB stars (Tokunaga 1997), diffuse interstellar medium (Onaka et al. 1996; Mattila et al. 1996), external star forming galaxies (Helou et al. 2000), remote ultraluminous infrared galaxies (ULIRGs) (Yan et al. 2005), and sub-millimeter galaxies even at $z=2.8$ (Lutz et al. 2005). They are supposed to be carried by small carbonaceous dust including polycyclic aromatic hydrocarbons (PAHs) and/or PAH-like species such as quenched carbonaceous composite (QCCs) (Leger & Puget 1984; Allamandola et al. 1985; Sakata et al. 1984). They are stochastically excited by absorbing a single ultraviolet (UV) photon and release the energy with a number of infrared photons in cascades via several lattice vibration modes of aromatic C-C and C-H (Allamandola et al.1989). Note that bulk QCC or amorphous carbon are not likely to be carriers of the UIR bands since the absorbed photon energy will not be confined within the aromatic group within/attached to the bulk QCC or amorphous carbon dust (Li & Draine 2002; Draine & Li 2006). The portion of the pumping energy to each vibration mode is supposed to be controlled by the physical and chemical conditions of the carriers such as the charging state, the molecular structure, and the size of the carriers, which follow as a consequence of the physical processing in the incident radiation environment as well as of the molecular evolution in chemically reactive regions. Therefore, understanding the systematic differences in UIR spectra in terms of the variation in the nature of the carriers in various astrophysical environments is, above all, important to use UIR bands as a useful probe of the local physical conditions (cf. Peeters et al. 2004).

The ionization state of PAHs is one of the most significant factors to affect the spectral characteristics of the UIR bands. The ionization of PAHs is controlled by $U/n_e$, where $U$ is the strength of the incident radiation field that acts in promoting the photo-ionization of PAHs and $n_e$ is the electron density that plays a role in the recombination. A large $U/n_e$ ratio favors positively ionized PAHs (Bakes et al. 2001). Past laboratory experiments and theoretical studies have shown that the UIR bands in the 6-9μm region are much weaker than those in the 11-14μm region when PAHs are neutral, however, that they become as strong as those in 11-14μm region when PAHs are ionized (de Frees et al. 1993; Szczepanski & Vala 1993; Bakes et al. 2001). Actually several studies report that the variations in the ratios of 7.7μm/11.2μm and/or 8.6μm/11.2μm bands, for example, within a reflection nebula along the distance from the ionizing source (Bregman & Temi 2005; Joblin et al. 1996), among the Herbig Ae/Be stars with different spectral types (Sloan et al.2005) and between the inner and outer Galactic plane (Sakon et al. 2004) have been reasonably explained by the changing in ionization status of the carriers of the UIR bands.

The variations of 11.0μm feature in the real astrophysical object have been firstly reported by Sloan et al. (1999). They observed the reflection nebula NGC1333 SVS 3 using the 5m Hale telescope at Palomar, and found an excess at 10.8-11.0μm and a feature around at 10μm increase relative to the 11.2μm feature in the close area to the illuminating early B star SVS 3. Recently, Werner et al. (2004) has reported the increase in 11.2μm /11.0μm with increasing distance from the central star based on the observation of reflection nebulae NGC7023 with Infrared Spectrograph (IRS) on the Spitzer Space Telescope. Bregman et al. (2005) have investigated the variation in the band ratio of 11.2μm/7.7μm within a reflection nebula along the distance from the central star and have made a quantitative evaluation of the relation between the 11.2μm/7.7μm and the ratio of the incident radiation field strength to the electron density. However, the 7.7μm and 11.2μm features come from different vibration modes (the former one corresponds to aromatic C-C stretching and the latter one to aromatic C-H out-of-plane bending, see Allamandola et al. 1989 for details), and their relative strengths are affected by various factors other than ionization, such as, the degree of hydrogenation, the molecular structures, and the molecular sizes (Peeters et al. 2002). On the other hand, the 11.0μm and 11.2μm feature come from the same vibration modes but with different ionization status of the carriers and, therefore, the 11.0μm/11.2μm band ratio can be used as more direct and quantitative measure for the ionization of PAHs than the 11.2μm/7.7μm band ratio.

In this work we aim to quantitatively evaluate the UIR 11.0μm/11.2μm band ratio in terms of the relation with *U/ne* by investigating the spectral changes in these features around the Herbig Ae/Be system MWC1080 using the Cooled Mid-Infrared Camera and Spectrometer (COMICS; Kataza et al. 2000) on board the 8m Subaru Telescope. Quite recently, Habart et al. (2006) have presented the spatially resolved PAH emission in the inner disks of nearby Herbig Ae/Be stars using adapted optics system on board NACO/VLT. Our attempts would surely be useful to understand the physical conditions of the disk around Herbig Ae/Be stars as well as the evolution of disk materials including carbonaceous dust when the spatially well-resolved mid-infrared spectra of nearby Herbig Ae/Be stars are obtained.

## 2. Observations and Data Reduction

### 2.1. *Target*

The mid-infrared spectroscopic data of MWC1080 were taken on the nights of 2005 July 16-17 (UT). MWC1080 is a non-isolated Herbig Ae/Be star located at the distance of 1.0kpc (Eisner et al. 2004) surrounded by a bright reflection nebulae (Herbig 1960) and the spectral type of the central star is classified as B0 (Choen & Kuhi 1979). Recently Wang et al. (2006) report the existence of at least 45 faint young low-mass stars within 0.3pc radius from the central star based on the observation of the CFHT with the high resolution adaptive optics. Optical bipolar outflows in the form of Herbig-Haro (HH) objects or HH-like jets with the radial velocity of 400 km/s have been discovered predominantly in the east of MWC1080 (Poetzel et al. 1992).

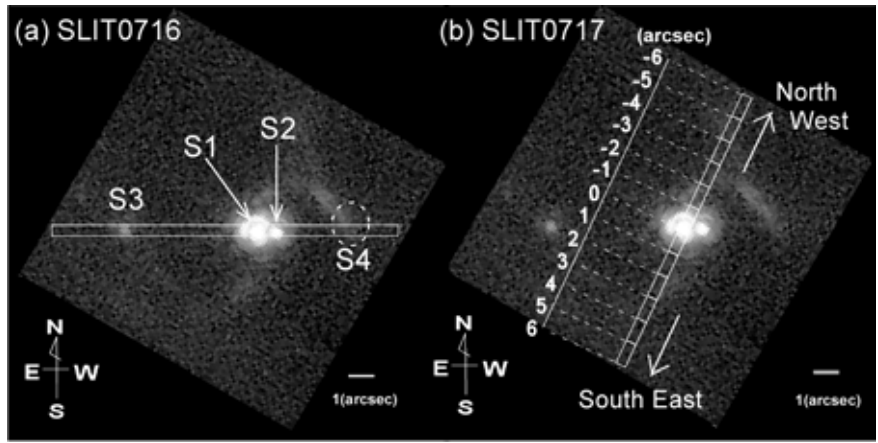

Figure 1.—Slit positions of (a) SLIT0716 and (b) SLIT0717 overlaid with the Subaru/COMICS N-band 11.7μm image of MWC1080.

We have obtained the N-band low-resolution spectra with the resolving power of R~250 using the 0".33 slit. We also carried out the imaging observation at 11.7μm ($\Delta\lambda=1.0$μm) to adjust the slit position. In order to cancel the high infrared background radiation, secondary mirror chopping was used at the frequency of 0.45 Hz with 20" throw. The spectra were obtained along two slit positions. One is set at a position angle of $-88^\circ.75$ so that it went across the

nebula(S4), the illuminating star(S1), the companion star(S2), and the nearby source(S3) (hereafter SLIT0716, see Fig.1a) and the total integration time was 810 sec, which enabled us to obtain the spectra of S1, S2, S3, and S4 with the signal-to-noise ratio larger than 10. The other one is set at a position angle of -25°.00 to observe the nebulae and the illuminating star (hereafter SLIT0717, see Fig.1b) and the total integration time was 1800 sec, which was sufficient enough to observe the variations in the diffuse UIR emission.

Non-isolated Herbig Ae/Be stars provide us decisively an ideal environment to investigate the ionization effect of PAHs on their spectra. Herbig Ae/Be stars are a pre-main sequence object of 2-8$M_\odot$ and the ionizing regions of hydrogen gas are restricted only within a few AU from the central star. The low $n_e$ environment, where carbon atoms instead of hydrogen dominantly supply electrons, realizes extremely high $U/n_e$ in the vicinity of the central star (within several hundred AU) compared to HII regions, taking account of the typical atomic abundance of C/H ~ $2\times10^{-4}$ for pre-main sequence objects (Snow & Witt 1995).

## 2.2. *Data Reduction*

The standard chopping subtraction and flat-fielding by thermal spectra of the telescope cell cover were employed. Then the spectra of MWC1080 along the SLIT0716 and SLIT0717 as well as the standard star HD3712 were obtained. The wavelength calibration was performed using the atmospheric emission lines and the uncertainty was estimated to be less than 0.0025μm (Okamoto et al. 2003). The spectra of MWC1080 along SLIT0716 and SLIT0717 were divided by the standard star HD3712 spectrum for the purpose of correcting the atmospheric absorption, and then we multiplied the resulting spectra by the template spectrum of the standard star provided by Cohen et al. (1999). Finally, we adjusted the flux measured in the N8.8 and N12.4μm imaging bands to correct the slit throughput. The uncertainty in the flux at each wavelength was estimated from the noise in the blank sky. We note that the ozone absorption increases the uncertainty in the 9.3-10.0μm region.

## 3. Results

### *3.1. Obtained Spectra along the SLIT0716*

Fig.2 shows the obtained spectra along SLIT0716. S1 is the illuminating central B0 star and the obtained spectrum is dominated by the strong emission from the photosphere. A slight dent in 9-10.5μm seems to be the effect of absorption by amorphous silicate but we note that this wavelength region is suffered by the atmospheric ozone absorption. On the other hand, S2 is the companion located at less than one arcsecond west to S1 and the obtained spectrum shows the clear presence of crystalline silicate features. The peaks around at 9.3μm and 10.5μm are supposed to be carried by crystalline enstatite, and those around at 11.3μm and 11.9μm are supposed to be carried by crystalline forsterite. The spectrum at S3 exhibits a feature peaking around 11.3μm, where the contribution from the UIR 11.2μm cannot be distinguished from that from the crystalline forsterite. However, we cannot further discuss the dust composition for this spectrum since other features characteristic to PAHs nor to crystalline forsterite are hardly recognized due to the relatively low signal-to-noise ratio. S4 is the diffuse nebula region and the obtained spectrum shows a series of the UIR bands, including those at 8.6μm and 11.2μm.

### *3.2. Changing in UIR solo-CH bond spectra along the SLIT0717*

PAHs in the diffuse nebulae of MWC1080 are supposed to be illuminated by the central B0 star and we assume that the projected distance from the central B0 star corresponds to the actual distance from the heating source. The spectrum of the diffuse nebula is actually dominated by the UIR features as observed in S4 on SLIT0716.

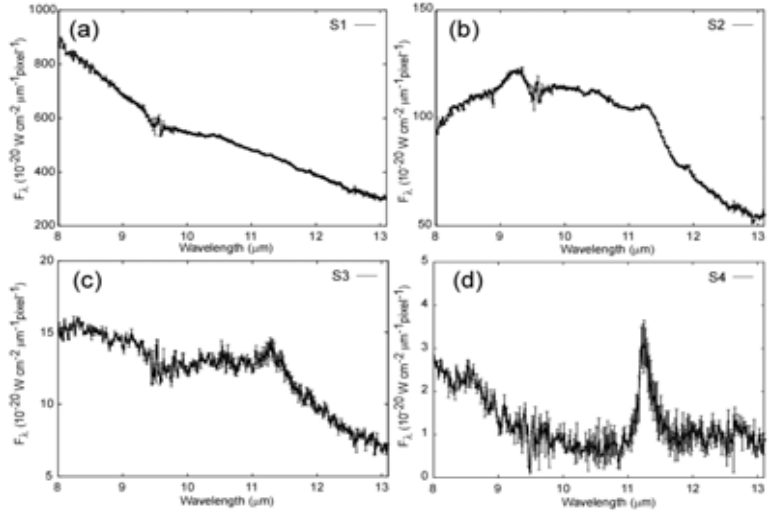

Figure 2. — Observed spectra at (a) S1, (b) S2, (c) S3, and (d) S4 along the SLIT0716. A pixel corresponds to an area of 0.165" x 0.165".

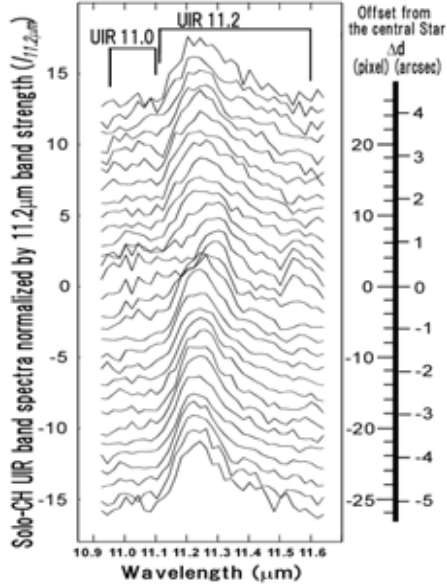

Figure 3. — Variations in the profiles of solo-CH bond features along SLIT0717. Each spectrum is normalized by the total intensity of the solo-CH bond features. The spectrum at the central star position corresponds to zero offset. The spectra are shifted

Along SLIT0717, therefore, we can investigate the spectra of PAHs in various strengths of the radiation field. In order to compare the profiles of these features at different positions, we subtract a local underlying continuum defined by linear interpolation between the average values around 10.9μm and 11.65μm (e.g. Witteborn et al. 1989). Figure 3 shows the variation in the continuum-subtracted UIR spectra of solo-CH modes normalized by the total intensity of their strengths along SLIT0717. We recognize a small hump located on the blue shoulder of the distinct UIR 11.2μm feature extending from 10.95μm to 11.1μm (hereafter the UIR 11.0μm band, see Fig. 3). In this analysis, the strength of the UIR 11.0μm band is defined by the integration of the continuum-subtracted emission from 10.95μm to 11.1μm, and that of the UIR 11.2μm band is defined by the integration of the continuum-subtracted emission from 11.1μm to 11.6μm. 1-$\sigma$ errors for each band strengths are defined by $(\sigma_m^2+\sigma_b^2)^{1/2}$, where $\sigma_m$ is the measurement error and $\sigma_b$ is the uncertainty in the baseline estimation. $\sigma_m$ is defined by $\sigma_m = \delta_{sky} \times \Delta\lambda$, where $\delta_{sky}$ is the standard deviation of flux density of the blank sky spectrum in units of W m$^{-2}$ μm$^{-2}$ pixel$^{-1}$ within the wavelengths used for calculating the strength of each UIR band and $\Delta\lambda$ is the width of each wavelength region. $\sigma_b$ is defined by $\sigma_b = \eta_{base} \times \Delta\lambda$, where $\eta_{base}$ is the standard deviation of flux density in units of W m$^{-2}$ μm$^{-2}$ pixel$^{-1}$ of the spectrum at each position along SLIT0717 within the wavelength range used for the continuum definition.

Figure.4*a* shows the spatial distribution of the intensities of UIR 11.0μm and 11.2μm features as a function of the offset from the central B0 star along SLIT0717. The UIR 11.0μm feature is significantly detected with generally better Signal-to-Noise (S/N) ratios than ~3 where the 11.2μm feature has local peaks (i.e. Δd ~ -4, -2, 0, and 3; see Fig.4) in its intensity distribution (Fig. 4a). Such regions are expected to have larger column density of PAHs and/or better supply of ultra-violet (UV) photons than other regions. Figure 4b shows the spatial variations of the relative band strength of the UIR 11.0μm to 11.2μm features along SLIT0717. We find the ratio increases up to ~0.3 in the vicinity of the central star. Among the above four regions around at Δd = -4, -2, 0, and 3, where the UIR 11.0μm feature is significantly detected in each spectrum, the nearest position to the central star (Δd ~ 0) shows the largest ratio while the most distant position from the central star (Δd ~ -4) shows the smallest ratio.

Taking account of the fact that the 11.0μm feature is assigned to a solo CH out-of-plane wagging mode of cationic PAHs while the 11.2μm feature to a solo CH out-of-plane bending mode of neutral PAHs (Hudgins & Allamandola 1999), the large ratio of the UIR 11.0μm/11.2μm in the vicinity of the central star can be interpreted as the promotion of PAHs' ionization to cationic species. We note that the UIR 11.2μm feature emitted from large PAHs may somewhat blue shifted in its peak position (Pathak & Rastogi 2006; Hudgins & Allamandola 1999) and a slight contribution from largest members of neutral PAHs to our calculated intensity of 11.0μm feature would be possible. However, in the following analysis we assume that our calculated intensity of 11.0μm feature is dominated by the emission carried by the CH out-of-plane wagging mode of cationic PAHs. We also note that some types of crystalline silicate (e.g. crystalline forsterite; Koike et al. 2003) can contribute to the spectra in 11μm region in the vicinity of the Herbig Ae/Be stars with a broad band feature peaking at 11.3μm as can be seen in the spectrum of S2 on SLIT0716 (see Fig.2b), but their band widths are typically broader than those of UIR features. Since our continuum subtracted spectra do not show an increase in the band width of 11.2μm feature in the vicinity of the central star (Fig. 3), we assume that spectra above the continuum in the 11μm region along SLIT0717 are dominated by UIR bands in the following analysis.

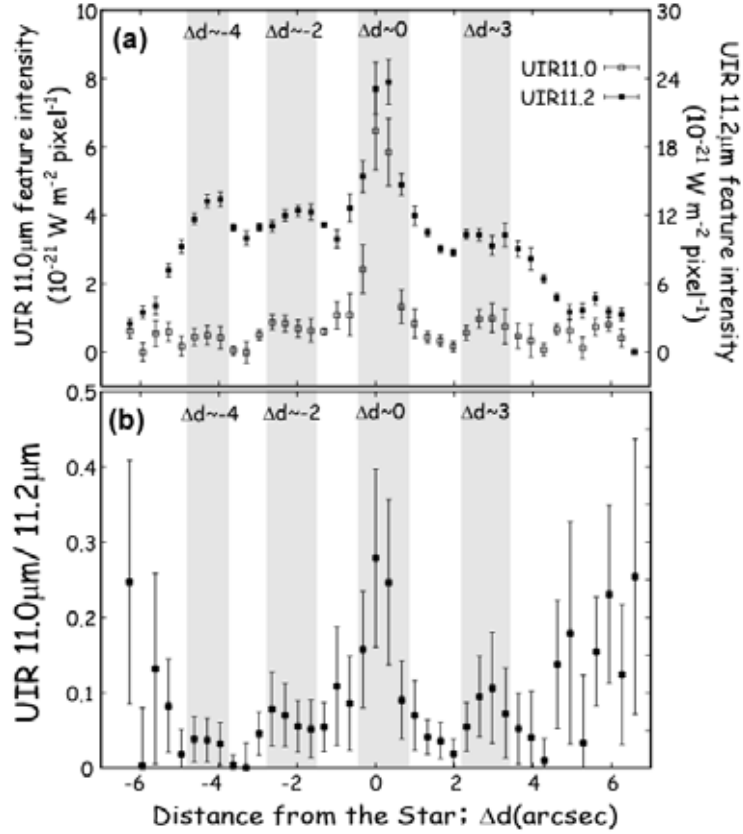

Figure 4. — Spatial variations a) in the UIR 11.0μm and 11.2μm intensities and b) in the relative band strengths of UIR 11.0μm to 11.2μm features as a function of the offset, Δd (arcsec), from the central B0 star along SLIT0717. A pixel corresponds to an area of 0.165" x 0.165". Regions where 11.2 μm feature has local peaks of in its intensity distribution (Δd ~ -4, -2, 0, and 3) are indicated with shadows.

## 4. Discussion

In this section we examine a quantitative relation between the band ratio of the UIR 11.0μm/11.2μm and the ratio of the interstellar radiation field strength to the electron density $U/n_e$ assuming a simple model in which the central B0 star is located inside in the spherically symmetric nebula. In this model, the interstellar radiation field strength at a distance *r* from the B0 star *U(r)* in units of the solar vicinity is given by,

$$U(r) = \frac{\int_{912\text{Å}}^{1\mu m} \pi B_\lambda(T_*) \left(\frac{R_*}{r}\right)^2 d\lambda}{\int_{912\text{Å}}^{1\mu m} 4\pi J_\lambda^\odot d\lambda}, \quad (1)$$

where $T_*=10^{4.31}$ K is the effective temperature, $R_*=3.2 \times 10^9$ m is the effective radius of the central B0 star (Millan-Gabet et al. 2001), and $J_\lambda^\odot$ is the interstellar radiation field of solar vicinity (Mezger et al. 1982).

We adopt a constant electron density $n_e=100^{+300}_{-50}$(cm$^{-3}$) from Poetzel et al. (1992) in the region in our analysis. We examine the relation between the UIR 11.0μm/11.2μm ratio and $U/n_e$ each in the north-west part of the slit and in the south east part of the slit (see Figure 5). We can clearly see a correlation between the UIR 11.0μm/11.2μm ratio and $U/n_e$, suggesting that the 11.0μm/11.2μm ratio can be a measure for the interstellar radiation field strength and the electron density. A relatively large scatter at small $U/n_e$ is supposed to originate from both the inhomogeneity in the electron density and the underestimation of the radiation field strength at the distant region from the central B0 star, where the contribution from faint low-mass stars (Wang et al. 2006) cannot be neglected.

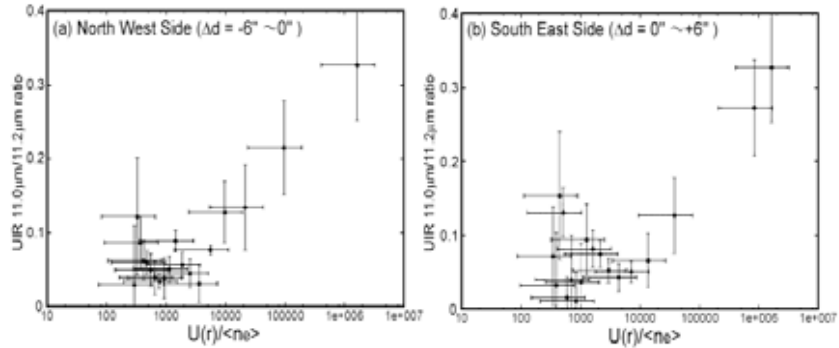

Figure 5. — (a)Relative band strengths of 11.0μm/11.2μm against $U(r)/n_e$ in the north west part of the slit (-6"<d<0") and (b) those in the south east part of the slit (0"<d<6"). The error bars for the horizontal axis are defined by the uncertainties in electron density $n_e=100^{+300}_{-50}$(cm$^{-3}$) and those for the vertical axis are calculated from the uncertainties in the estimation of the UIR band strengths.

Detailed modeling of the interstellar radiation field strength and the electron density as well as the additional spectroscopic observation of the UIR 11.0 and 11.2μm around spatially resolved non-isolated Herbig Ae/Be objects are quite important to examine the quantitative and robust relation between the UIR 11.0μm/11.2μm ratio and $U/n_e$.

## 5. Summary

We present mid-infrared N-band spectroscopy of the Herbig Ae/Be system MWC1080 using the Cooled Mid-Infrared Camera and Spectrometer (COMICS) on board the 8m Subaru Telescope. MWC1080 has a geometry such that diffuse nebulous structures extend around the central Herbig B0 type star. We focus on the properties of polycyclic aromatic hydrocarbons (PAHs) and PAH-like species, which are thought to be the carriers of the unidentified infrared (UIR) bands in such environments. A series of the UIR bands at 8.6, 11.0, 11.2, and 12.7μm are detected throughout the system and we find a clear increase in the UIR 11.0/11.2μm ratio in the vicinity of the central star. Since the UIR 11.0μm feature is attributed to a solo-CH out-of-plane wagging mode of cationic PAHs while the UIR 11.2μm feature to a solo-CH out-of-plane bending mode of neutral PAHs, the large 11.0/11.2μm ratio directly indicates a promotion of the cationic ionization of PAHs. This work suggests an application and robust use of the UIR 11.0/11.2μm ratio as a valid probe of the local interstellar radiation field strength and the electron density.

**Acknowledgments**

The authors are grateful to all the staff members of the Subaru Telescope for the continuous support. I.S. especially thanks Drs. Hideko Nomura, Amit Pathak, and Hiroshi Kimura for useful comments and discussion. This work is supported by a Grant-in-Aid for the Japan Society for the Promotion of Science (JSPS). Y.K.O. is supported by a Grant-in-Aid for young scientists (#17740103) by the Ministry of Education, Culture, Sports, Science and Technology, Japan.

**References**

Allamandola, L. J., Tielens, A. G. G. M. & Barker, J. R. 1985, ApJ, 290, 25
Allamandola, L. J., Tielens, A. G. G. M. & Barker. J. R. 1989, ApJS, 71, 733
Bakes, E. L. O., Tielens, A. G. G. M., Bauschlicher, C. W. ApJ, 556, 501
Bregman, J. & Temi, P. 2005, 621, 831
Cohen, M. & Kuhi, L.V. 1979, ApJS, 41, 743
Cohen, M., et al. 1999, AJ, 117, 1864
Draine,B.T. & Li, A. 2006, ApJ, in press [arXiv: astro-ph/0012546]
de Frees, D. J. et al. 1993, ApJ, 408, 530
Eisner, J. A., 2004, ApJ, 613, 1049
Habart,E., Natta, A., Testi, L., & Carbillet, M. 2006, A&A, 449, 1067
Helou, G., et al. 2000, ApJ, 532, 21
Hudgins, D. M. and Allamandola, L.J., 1999, ApJ, 516, L41
Joblin,C., Tielens, A.G.G.M.,Geballe,T.R.&Wooden,D.H. 1996, ApJ, 460, L119
Kataza, H. et al. 2000, in Proc. of the SPIE Conf. Ser, 4008, 1144
Koike, C., et al. 2003, A&A, 399, 1101
Li, A. & Draine, B.T., 2002, ApJ, 572, 232
Lutz, D. et al. 2005, ApJ, 625, L83
Leger, A. & Puget, J. L. 1984, A&A, 137, L5
Mattila, K. et al. 1996, A&A, 315, L5
Millan-Gabet,R., et al. 2001, ApJ, 546, 358
Okamoto, Y. K. et al. 2003, in Proc. of the SPIE Conf. Ser. 4841, 169
Onaka, T.,Yamamura,I.,Tanabe,T., Roellig,T.L.,& Yuen,L.1996, PASJ, 48, L59
Pathak, A. & Rastogi, S. 2006, Proc.of 36[th] COSPER Scientific Assembly, p.432
Peeters, E., Spoon, H. W. W., Tielens, A. G. G. M. 2004, ApJ, 613, 986
Peeters, E. et al.. 2002, A&A, 390, 1089
Poetzel,R., et al. 1992, A&A, 262, 229
Sakata, A., Wada, S., Tanabe, T., Onaka, T. 1984, ApJ, 287, 51
Sakon, I. et al. 2004, ApJ, 609, 203 (erratum 625, 1062 [2005] )
Sloan, G. C. et al. 1999, ApJ, 513, 65
Sloan, G. C. et al. 2005, ApJ, 632, 956
Snow, T. P., & Witt, A. N. 1995, Science, 270, 1455
Szczepanski, J. & Vala, M. 1993, ApJ, 414, 646
Tokunaga, A.T. 1997, ASP Conf. Ser. 124, 149
Wang, S., Looney, L. W., Brandner, W., Close, L. M. 2006, IAUS, 237, 238
Werner, M.W., et al. 2004, ApJS, 154, 309
Witteborn, F. C., et al. 1989, ApJ, 341, 270
Yan, L. et al. 2005, ApJ, 628, 604